\documentstyle[12pt,aasms4]{article}
\slugcomment{}

\def\bm{{\mu\hskip-6.4pt\mu}}
\def\bv{{\bf v}}
\def\cl{{\rm cl}}
\def\bw{{\rm f}}
\def\obs{{\rm obs}}
\newcommand{\be}{\begin{equation}}
\newcommand{\ee}{\end{equation}}
\journalid{}{}
\articleid{}{}

\begin{document}

\title{The Proper Motion of NGC 6522 in Baade's Window\footnote{Based
in part on data obtained at the Anglo--Australian Observatory and at
the Cerro Tololo Inter--American Observatory, NOAO, which is operated
by the Association of Universities for Research in Astronomy, Inc.
(AURA), under cooperative agreement with the National Science
Foundation.}}

\author{Donald M. Terndrup, Piotr Popowski, and Andrew Gould\altaffilmark{2}}
\affil{Department of Astronomy, The Ohio State University, \\
174 W. 18th Ave., Columbus, Ohio 43210 \\
Electronic mail: (terndrup,popowski,gould)@astronomy.ohio-state.edu}

\author{R. Michael Rich}
\affil{Department of Astronomy, Pupin Laboratories, Columbia University, 
538 West 120th Street, New York, New York  10027 \\
Electronic mail: rmr@astro.columbia.edu}

\author{Elaine M. Sadler}
\affil{School of Physics, University of Sydney,
NSW 2006, Australia \\
Electronic mail: ems@physics.usyd.edu.au}

\altaffiltext{2}{Alfred P. Sloan Fellow.}

\begin{abstract}
We have detected seven stars with a common proper motion which are
located within $2\farcm 5$ of the globular cluster NGC 6522 in the
Baade's Window field of the Galactic bulge.  We argue that these stars
are members of the cluster, and derive a weighted mean proper motion
and heliocentric radial velocity of $\bar\mu_\ell = 1.4 \pm 0.2$ mas
yr$^{-1}$, $\bar\mu_b = -6.2 \pm 0.2$ mas yr$^{-1}$, $\bar v = -28.5
\pm 6.5$ km s$^{-1}$. We rederive the distance to NGC 6522 ($0.91 \pm
0.04 R_0$, where $R_0$ is the Galactocentric distance) and metallicity
(${\rm [Fe/H]} = -1.28 \pm 0.12$), making use of recent revisions in
the foreground extinction toward the cluster ($A_V = 1.42 \pm 0.05$).
We find the spatial velocity of the cluster and conclude that the
cluster stays close to the Galactic center, and may have experienced
significant bulge/disk shocking during its lifetime.
\end{abstract} 

\keywords{astrometry -- Galaxy: abundances --
globular clusters: individual (NGC 6522)
-- stars: kinematics}

\section{Introduction}

Three-dimensional space motions of individual stars and stellar systems
throughout the Galaxy are an important probe of the Milky Way's
gravitational potential and of the kinematics of different populations
within it.  Most of what we know, for example, about the disk and halo
of the Galaxy has come from studies of the radial velocities and proper
motions of stars in the local volume of space near the Sun.  In recent
years, this has been steadily extended to samples throughout the
Galaxy, particularly for globular clusters (\cite{cud93}; \cite{dau96})
and the Large Magellanic Cloud and other Galactic satellites (e.g.,
\cite{jon94}; \cite{sch96}; \cite{iba97}).

A potentially important sample for {\it in situ} studies of the
Galactic bulge is the proper motion survey of Spaenhauer et al.\ (1992
- hereafter SJW).  This survey contains relative proper motions of 427
stars in the Baade's Window field of the nuclear bulge at $(\ell, b) =
(1.0^\circ, -3.9^\circ)$, which samples stellar orbits relatively close
to the Galactic center (minimum line of sight distance $\approx 550$
pc).  The proper motions presented by SJW are with respect to
the mean motion of the entire sample (i.e., they set $\bar\mu_\ell
\equiv 0,  \bar\mu_b \equiv 0$), since no extragalactic reference frame
is available.

We have been conducting an extensive photometric and spectroscopic
followup of the SJW survey.  In our first paper (\cite {tsr95}), we
presented methods of analysis and demonstrated that the sample contains
a considerable number of stars located in front of the bulge.  In the
second paper (\cite{srt96}), we derived individual metallicities and
distances to the SJW stars.  In preparing for a full analysis of the
space motions in Baade's Window (\cite{ric98}), we have discovered that
a number of stars with proper motions in Baade's Window are likely to
be members of the globular cluster NGC 6522.  In \S 2 we describe the
procedure by which the stars are identified as cluster members, measure
the proper motion of the cluster relative to the mean proper motion of
the SJW sample, and estimate the cluster's radial velocity.  In \S 3,
we analyze the cluster color-magnitude diagram (CMD) and derive
estimates for the cluster metallicity and distance.  In \S 4, we
estimate the space motion of the cluster and discuss its orbit.

\section{Identification of Cluster Members}

We begin by selecting the 406 stars that have radial velocities
(\cite{tsr95}) out of the 427 stars\footnote{SJW tabulated 429 stars,
but listed the stars Arp 3-144 and 4-027 twice.} in the SJW survey.  We
first demonstrate that members of NGC 6522 are present in the SJW
survey by noting that there are several stars with a common proper
motion which have radial velocities near the value previously
determined for the cluster.

The radial velocity of NGC 6522 has been reported several times in the
literature.  \cite{ric90} obtained the radial velocity of two stars
near the cluster center from spectra of resolution 1.9\AA; the average
heliocentric velocity of these was $v = -25.6 \pm 1.5$ km s$^{-1}$,
identical to the value of $-25 \pm 16$ km s$^{-1}$ derived from
Fabry-Perot observations at H$\alpha$ by \cite{smith76}.  Averaging a
variety of observations, including the Smith et al. (1976) value,
\cite{web81} obtained $v = +8 \pm 16$ km s$^{-1}$.  As Webbink himself
noted, however, the velocities he included in the average for NGC 6522
formed a bimodal distribution;  it is therefore likely that several
bulge stars were included in the computation. In his compilation of
globular cluster velocities, \cite{zin83} repeats the \cite{web81}
value. Pryor \& Meylan (1993) quote the radial velocity as $-10.4 \pm
1.5$ km s$^{-1}$ from an apparently unpublished paper.  Recently,
\cite{rut97} obtained \ion{Ca}{2} spectra of 18 stars near the cluster,
and derived $v = -18.3 \pm 9.3$ km s$^{-1}$, where the uncertainty is
the standard error of the mean.

In the upper panel of Figure 1, we display the proper motion vector
point diagram for the complete SJW survey, while in the lower panel we
show those stars located within $2\farcm 5$ of the cluster center with
radial velocities statistically consistent at the
$1\sigma$ level with the Smith et al.\ (1976) value of $-25 \pm 16$ km
s$^{-1}$.  The error bars in the lower left hand corner of Figure 1
show the mean error in proper motion for the stars in the sample.  In
the lower panel, one clumping of points is present at $(\mu_\ell,\mu_b)
\sim (1.5,-6)$ mas yr$^{-1}$.  As we will show, this is the most likely
proper motion of the cluster because stars near this clump lie on the
giant branch for NGC 6522 and have absorption line strengths unlike
most of the stars in the bulge;  other possible clumps of stars in
Figure 1 do not yield this {\it a posteriori} consistency with the
cluster CMD and metallicity.

To identify cluster members, we employ an iterative scheme which
searches for stars in the SJW sample that are simultaneously within
$1.5 \sigma$ of the mean cluster motion $(\bar\mu_{\ell,c},
\bar\mu_{b,c}, \bar v_c)$ in each proper motion component and in radial
velocity.  For example, for heliocentric radial velocity component,
we demand that
\be
     \mid v_i - \bar v_c \mid \leq 1.5\sigma_i,
\ee
where $v_i$ is the velocity of an individual star,
$\bar v_c$ is the mean velocity of cluster members, 
\be
\sigma^2_i \equiv \epsilon_i^2 + \sigma^2_c(v),
\ee
$\epsilon_i$ is the measurement error, and $\sigma_c(v)$ is the
estimated scatter due to the internal velocity dispersion of the
cluster.  Similar expressions are used for $\mu_\ell$ and $\mu_b$.  We
adopt $\sigma_c(v) = 6.7$ km s$^{-1}$ (\cite{pry93}).  For a distance
to NGC 6522 of $7.1$ kpc (below), this corresponds to
$\sigma_c(\mu_\ell) = \sigma_c(\mu_b) = 0.20$ mas yr$^{-1}$, which we
adopt for the dispersion in each component of the proper motion.  (The
results of this procedure are not very sensitive to the choice of
$\sigma_c$ because $\epsilon_i \geq \sigma_c$ for most of the sample.)

We began the iteration with a few stars near the clump in the lower
panel of Figure 1.  Then at each step, the list of stars near the mean
cluster motion was used to recompute the weighted means
$(\bar\mu_{\ell,c}, \bar\mu_{b,c}, \bar v_c)$, with weights
$1/\sigma_i^2$ in each component.  This process quickly converged on
seven stars, which are listed in Table 1.  The star names in the first
column are from \cite{arp65}.  Columns 2-7 show the radial velocity and
photometry (\cite{tsr95}).  The next four columns contain the proper
motion and errors from SJW, where the units are mas yr$^{-1}$.  The
last column shows the angular distance of each star in arcmin from the
center of NGC 6522.

At this point we examined the errors in proper motion and radial
velocity to confirm that they were estimated correctly, for if the
selected stars truly are members of the cluster the scatter in
projected motion should be almost entirely due to observational error.
Via a $\chi^2$ test, we determined that the scatter in proper motions
about the mean value is consistent with the errors, but the errors for
the radial velocities were too large by a factor of $\sim 1.7$.  The
reason for this is not completely clear.  As discussed by Terndrup et
al.\ (1995), the radial velocity errors were estimated from the width
of the peak in the cross-correlation between the bulge stars and a set
of radial velocity templates;  intercomparisons between multiple
measures allowed them to confirm that the error estimates were, in
general, approximately right.  The possible cluster members, however,
happen to be among the most metal-weak of the stars in the SJW sample
(below), which could have led to an overestimate of the error because
of the resulting mismatch between these stars and the velocity
standards.  An examination of the estimated errors as a function of
metallicity (\cite{srt96}) reveals that this explanation is plausible:
the mean estimated error in radial velocity for stars with [Fe/H] $<
-0.7$ is 25 km s$^{-1}$, compared to 15 km s$^{-1}$ for stars with
[Fe/H] $>-0.7$.

We then scaled the radial velocity errors downward by a factor of 1.7
(but not below a reasonable minimum value of 7 km s$^{-1}$) and repeated
the iterations, obtaining the same list of candidate members of NGC
6522.  The weighted mean cluster motion and standard errors of the mean
from this sample are
\begin{eqnarray}
\bar\mu_{\ell,c} & = &  +1.2 \pm 0.2 \ {\rm mas\ yr}^{-1}, \nonumber \\
\bar\mu_{b,c}    & = &  -6.0 \pm 0.2 \ {\rm mas\ yr}^{-1},  \\
\bar v_c         & = & -28.5 \pm 6.5 \ {\rm km\ s}^{-1}. \nonumber 
\end{eqnarray}
The radial velocity we derive is in statistical agreement with the
value of $-18.3 \pm 9.3$ km s$^{-1}$ from spectra of 18 stars near the
cluster reported by Rutledge et al.\ (1997).

Finally, in order to explore the sensitivity of our selection method,
we compute membership probabilities for the SJW stars with radial
velocities assuming a Gaussian distribution for the cluster and field
(mostly the bulge) and taking into account the errors in the individual
measurements (cf.  \cite{din96}).  We write the mean apparent motion of
the field as $\lbrace\bar \mu_{\ell,f}, \bar \mu_{b,f}, \bar
v_f\rbrace$ and the dispersion as $\lbrace\sigma_f(\mu_\ell),
\sigma_f(\mu_b), \sigma_f(v)\rbrace$.  Denoting with the subscript $c$
the equivalent quantities for the cluster, determined above, we define
the membership probability for star $i$ as
\be
P_i  = { {\rho_{i,c}}
               \over
               {\rho_{i,c} + \rho_{i,f}}
             },
\ee
where the densities $\rho_{i,c}$ and $\rho_{i,f}$ of cluster and field
stars are represented as the Gaussian distributions
\begin{eqnarray}
\rho_{i,c} &=& {\frac
                  {N_c}{ 
                           (2 \pi)^{3/2}
			   \sigma_i(\mu_\ell) \sigma_i(\mu_b) \sigma_i(v) 
                              }
                  } 
                  \exp\left[ 
                      - { {(\mu_{\ell,i} - \bar\mu_{\ell,c})^2} 
                               \over {2\sigma^2_i(\mu_\ell)}}
                      - { {(\mu_{b,i} - \bar\mu_{b,c})^2} 
                               \over {2\sigma^2_i(\mu_b)}}
                       - { {(v_i - \bar v_c)^2} 
                               \over {2\sigma^2_i(v)} }
	          \right],
            \\
\rho_{i,f} &=&  {\frac
                  {N_f}{ 
                           (2 \pi)^{3/2}
			   \sigma_f(\mu_\ell) \sigma_f(\mu_b) \sigma_f(v) 
                        }
                  } 
                  \exp\left[ 
                      - { {(\mu_{\ell,i} - \bar\mu_{\ell,f})^2} 
                               \over {2\sigma^2_f(\mu_\ell)}}
                      - { {(\mu_{b,i} - \bar\mu_{b,f})^2} 
                               \over {2\sigma^2_f(\mu_b)}}
                       - { {(v_i - \bar v_f)^2} 
                               \over {2\sigma^2_f(v)} }
	          \right].
\end{eqnarray}
Note that in the expression for the cluster density function we
combined the errors of the individual stars and the adopted cluster
dispersion as in equation (2).  The values for the field were derived
from the SJW stars with velocities excluding the seven stars in Table
1, and were
$\bar\mu_{l,f} = 0.04$ mas yr$^{-1}$, 
$\sigma_f(\mu_\ell) = 3.1$ mas yr$^{-1}$,
$\bar\mu_{b,f} = 0.18$ mas yr$^{-1}$, 
$\sigma_f(\mu_b) = 2.7$ mas yr$^{-1}$,
$\bar v_f = -2.1$ km s$^{-1}$,
$\sigma_f(v) = 105$ km s$^{-1}$.  
As opposed to the cluster members, the field stars' dispersions are
dominated by the intrinsic velocity dispersions and not observational errors.
Figure 2 displays the membership
probabilities, expressed as percentages, against (top panel) the
distance from the center of NGC 6522 and (bottom panel) the $V$
magnitude.  The stars in Table 1 are those seven stars with the highest
membership probabilities;  all seven have $P \geq 85\%$.  There are only
three other stars with $P \geq 10\%$, and most of the sample has $P$
very nearly zero.  (The SJW sample contains almost no stars within
$1\arcmin$ of the cluster because this region was deliberately avoided
in their survey.)

\section{The distance to NGC 6522}

To support our claim that we have found members of NGC 6522 in the SJW
survey, we now examine photometry and line-strength indices
(\cite{tsr95}) for the stars in Table 1.  We also derive the distance
to NGC 6522, needed for the conversion of proper motion into a space
velocity, from its CMD and by direct comparison of its HB magnitude to
that of RR Lyrae stars in Baade's Window.  In this process, we obtain a
photometric metallicity estimate for comparison to recent results from
spectroscopy.

The distance to NGC 6522 is not known accurately:  although
there are several RR Lyrae stars near the cluster, it is difficult to
determine which are members and which are in the bulge (\cite{bla84};
\cite{wal86}; \cite{wal91}; Carney et al. 1995).  Consequently the only
distance estimates have come from analysis of the cluster's  CMD, and
these are sensitive to the assumed metallicity of and extinction
towards the cluster.  For example, \cite{tw94} derived [Fe/H] $= -1.6
\pm 0.2$ and $(m - M)_0 = 14.8 \pm 0.3$ ($d = 9.1^{+1.4}_{-1.2}$ kpc)
from comparison of the cluster CMD to the giant branches of globular
clusters in \cite{da90}.  This derivation was for $E(V - I) = 0.65 \pm
0.07$, and the \cite{da90} sequences were tied to the \cite{ldz90}
calibration of RR Lyrae absolute magnitudes of $M_V({\rm RR}) = 0.82 +
0.17{\rm [Fe/H]}$.  The metallicity that Terndrup \& Walker (1994)
derived was in agreement within the errors to the \cite{zw84} value
[Fe/H] $= -1.44$ from integrated colors.

Nearly simultaneously with \cite{tw94}, \cite{bar94} presented their
own CMD of the cluster and derived extinction values and metallicity by
comparison to the CMD of NGC 6752, which has [Fe/H] $ = -1.54$ (DaCosta
\& Armandroff 1990 and references therein).  They find $E(B - V) = 0.55
\pm 0.05$, which corresponds to $E(V - I) = 0.68 \pm 0.06$ (i.e., close
to the Terndrup \& Walker 1994 value) and a distance modulus of $(m -
M)_0 = 13.96$ ($d = 6.2$ kpc).  They assume a horizontal branch
luminosity of $M_V = +0.6$ and obtain a very approximate metallicity
estimate of ${\rm [Fe/H]} \approx -1.0$.  Together, these two estimates
imply that they were using a horizontal branch luminosity scale which
is slightly (0.05 mag) more luminous than the Lee et al.\ (1990) scale
employed by Terndrup \& Walker (1994).   Note, however, that the
difference in horizontal branch luminosity scale is far too small to
explain the 0.8 mag difference in the distance determinations;  the
largest effect comes from the choice of horizontal branch magnitudes on
the CMD in the two analyses.

To derive the metallicity of the cluster, we reanalyze the photometry
of Terndrup \& Walker (1994) using the method outlined by
\cite{sar94}.  Their method uses polynomial fits to the DeCosta \&
Armandroff (1990) giant branch sequences to measure simultaneously the
metallicity of and extinction toward a cluster from color-magnitude
diagrams in $V, V - I$.  In using this method, we make two changes to
reflect recent developments.  First, we use the \cite{sta96} reddening
map for Baade's Window, which allows us to correct the photometry of
the cluster (and of the probable cluster members in Table 1) for the
considerable variation in extinction with position.  Second, we adopt
the \cite{car92} calibration of the RR Lyrae luminosity ($M_V = 1.01 +
0.16\rm [Fe/H]$) instead of the Lee et al.\  (1990) scale implicit in
the Sarajedini (1994) approach;  since the metallicity slope is almost
the same for both calibrations, this has the effect of reducing the
derived distance by 0.2 mag.  We adopt the Carney et al.\ (1992) scale
so that our distance for NGC 6522 is on the same system as Galactic
center distance from RR Lyrae stars in Baade's Window (Carney et
al.\ 1995).  This way the relative distance between the cluster and the
Galactic center can be found independent of the zero point in the
calibration of RR Lyrae luminosity.

In employing the Stanek (1996) map, we first add an offset of $-0.10$
to $A_V$ (i.e., the reddening is less than in Stanek's map) as
discussed by \cite{gould98} and confirmed by \cite{alc98}.  Next, we
interpolate across a region of radius $2\arcmin$ around NGC 6522 which
is not included in the map.  An inspection of photographs of Baade's
Window (e.g., \cite{bla84}) shows that, at least near the cluster, the
gradient in extinction is primarily north-south with the extinction
lower to the north of the cluster.  In Figure 3, we show the Stanek
reddenings (with the corrected zero point) for the stars in the SJW
survey as a function of the declination difference between each star
and the cluster center (positive values indicate stars north of the
cluster center).  The straight line shows a linear fit to the
reddenings;  the rms scatter about this line is 0.06 in $A_V$.  From
this plot and the Stanek (1996) relation $A_V = 2.49 E(V - I)$, we
conclude that the total and selective extinctions towards the center of
the cluster are $A_V = 1.42 \pm 0.05$, and $E(V - I) = 0.57 \pm 0.02$.

In Figure 4 we display (left panel) a CMD in $V, V - I$ for stars
within $1\farcm 5$ of NGC 6522 (small points) along with the photometry
for the cluster members in Table 1 (large filled points). The right
panel shows the photometry for stars farther than $2\farcm 5$ from the
cluster center.  In this figure, the photometry is from the sources
listed in \cite{tw94} and Terndrup et al. (1995); all the colors and
magnitudes have been corrected to the reddening adopted for the cluster
center using the linear fit to the reddening with position in Figure
3.  The curved line is a quadratic fit to giant-branch stars on the
left panel of Figure 4 (not only the possible proper motion members)
using a $3\sigma$ rejection criterion.  Note that, in support of our
selection method,  the possible cluster members are near the cluster's
giant branch and well away from the majority of bulge stars.  (The
scatter about mean giant branch is significantly worse without the use
of the Stanek (1996) reddening map.)

Returning to the distance determination for the cluster, we note that
the level of the horizontal branch at a representative color for RR
Lyrae stars ($B - V$ = 0.80, at the reddening to the cluster) is $V =
16.50 \pm 0.15$.  This is consistent with the HB level on the much
better CMD of NGC 6522 from HST in \cite{sos97}; interpolating from
their CMD, we find $V_{\rm HB} = 16.52 \pm 0.07$.  Using this value for
$V_{\rm HB}$, we apply the Sarajedini (1994) method and derive [Fe/H]
$= -1.28 \mp 0.12$.  The metallicity we derive agrees with the Rutledge
et al.\ (1997) value [Fe/H] $= -1.21 \pm 0.04$, derived from infrared
\ion{Ca}{2}-triplet measures transformed to the high-resolution
abundance scale of \cite{cg97}.  The method simultaneously returns the
estimate of the cluster reddening of $E(V - I) = 0.58 \pm 0.03$ in
excellent agreement with the value $E(V - I) = 0.57 \pm 0.02$ derived
above from the corrected Stanek (1996) extinction map, and an estimate
of the cluster distance modulus of $14.3 \pm 0.1$ on the Carney et
al.\ scale (see also below).

In Figure 5, we plot the Lick Mg$_2$ indices as a function of color for
the Baade's Window sample (Terndrup et al.\ 1995).  The filled points
with error bars are for the stars in Table 1.  With the possible
exception of Arp 2-069, which has a value of Mg$_2$ about 1.8$\sigma$
above that of the others, all the possible cluster members have low
Mg$_2$ indices at a given color, unlike those for the bulge stars in
the SJW sample.  Since the calibration of the Lick indices with
metallicity is uncertain for stars of low metallicity (Sadler et al.
1996), we do not attempt to measure the abundance of NGC 6522 from
these Mg$_2$ indices -- we simply note that the probable members are
among the lowest-metallicity stars in Baade's Window, consistent with
both the photometric and spectroscopic determinations of the abundance
of NGC 6522.

A direct calculation of the relative distance between the cluster and
the Galactic center is as follows:  an analysis of the magnitudes of RR
Lyrae stars in Baade's Window shows that the dereddened visual
magnitude of the horizontal branch at typical colors of RR Lyrae stars
is $V_{0,{\rm BW}} = 15.33 \pm 0.03$ (\cite{alc98b});  there is an
additional error from the uncertainty in extinction which does not
enter here because we are comparing the cluster HB and the bulge HB in
the same field.  Taking the dependence of the horizontal branch level
as
\be
M_{V,{\rm HB}} = a + b{\rm [Fe/H]},
\ee
and using the definition of distance modulus
we write the difference in distance moduli between the cluster and the
Galactic center as
\be
{\Delta}_{c,BW} = 
 (V_{{\rm HB},c} - A_V - M_{V,c}) - (V_{0,{\rm BW}} - M_{V, {\rm BW}}),
\ee
where the subscript $c$ refers to the cluster and the subscript $BW$
to the RR Lyrae stars in Baade's Window.  With the adopted reddening to
NGC 6522, this becomes
\be
{\Delta}_{c,BW}
= (-0.23 \pm 0.08) 
                     + b({\rm [Fe/H]}_{\rm BW} - {\rm [Fe/H]}_c).
\ee
We found ${\rm [Fe/H]}_c = -1.28 \pm 0.12$ (above), and we adopt
${\rm [Fe/H]}_{\rm BW} = -1.1$ (\cite{wal91}).  Setting $b = 0.16 \pm
0.10$, we have
\be
{\Delta}_{c,BW} = -0.20 \pm 0.08.
\ee
For a distance to the Galactic center of 7.8 kpc, the cluster is
therefore at a distance of $7.1 \pm 0.3$ kpc.   We will adopt
this distance in the analysis that follows.

\section{Space Motion}

Because the SJW survey was not tied to an extragalactic reference
frame, some care is required to transform this measurement into an
estimate of the cluster's space motion.  We begin by writing the
observed vector proper motion $\bm_{\rm obs}$ as
\be
\bm_\obs = {\bv_{\cl,\perp} - \bv_{\odot,\perp}\over d_\cl}
- {\bv_{\bw,\perp} - \bv_{\odot,\perp}\over d_\bw} - \Delta \bm,
\ee
where $d_\cl$ and $\bv_{\cl,\perp}$ are the distance to the cluster and
its velocity transverse to the line sight, $d_\bw$ and
$\bv_{\bw,\perp}$ are the mean distance and transverse velocity of
bulge field stars in Baade's Window, $\bv_{\odot,\perp}$ is the velocity of
the Sun transverse to the line of sight, and $\Delta \bm$ is a
correction to be discussed below.  
(Note that $1 \, \rm mas\, yr^{-1}= 4.74 \,km\,s^{-1} \, kpc^{-1}$.)
Next, we solve for $\bv_{\cl,\perp}$
and write the result in a way that allows easy identification of the
role of various uncertainties,
\be
\bv_{\cl,\perp} = d_\cl\biggl(\bm_\obs + \Delta \bm + 
{\bv_{\bw,\perp}\over d_\bw}\biggr)
 + \bv_{\odot,\perp}\biggl(1 - {d_\cl\over d_\bw}\biggr).
\ee

We now estimate $\Delta\bm$, the mean proper motion of bulge stars
relative to the proper-motion frame of the SJW stars.  There are two
distinct sources of contamination that lead to an offset:  foreground
disk stars and the cluster stars themselves.  To estimate the offset
due to foreground disk stars, we focus on the subsample of 310 stars
(out of a total of 427) with spectroscopic distance estimates from
Sadler et al.\ (1996).  We compare the 241 stars with distances
$ > 4$ kpc with the full subsample of 310 stars and find an offset $(\Delta
\mu_\ell,\Delta \mu_b) = (0.16,-0.05)$ mas yr$^{-1}$.  
We assume that this offset is representative of all 427 stars.
The full SJW sample contains $\sim 330$ stars at distances $> 4$ kpc, of which 
$\sim 7$ are cluster members. The offset due to cluster stars is therefore
$7/330 \sim 2\%$ of the value given in equation (3), or 
$(\Delta \mu_\ell,\Delta \mu_b) = (0.02,-0.12)$ mas yr$^{-1}$.
Within the accuracy of our determinations, the combined offset from these two 
sources of bias is therefore
\be
(\Delta \mu_\ell,\Delta \mu_b) = (0.2,-0.2)\,{\rm mas\,yr^{-1}}.
\ee

The distance to the Galactic center is uncertain (e.g.\ Reid 1993), but
for present purposes, we treat it as being fixed at the distance given
by the Carney et al.\ (1992, 1995) scale, 7.8 kpc, and scale all
results to this value.  Recall that we have also estimated the distance
to the cluster on this scale.  Since Baade's Window is very close to
the minor axis of the bulge (at $\ell = +1^\circ$), a fair sample of
bulge stars should have $\bv_{\bw,\perp}=0$.  However, there are two
conflicting biases that can cause a deviation from this expected
value.  First, the underlying SJW sample is biased toward brighter
stars and hence toward stars on the near side of the bulge. Second, the
cone of observation is wider on the far side of the bulge than the near
side and so contains more far-side stars.  By examining the
distribution of spectroscopic distances, we estimate that the median of
the distribution lies $0.3\pm 0.3\,$kpc behind the peak which we
identify with the galactocentric distance.  The SJW sample is therefore
slightly biased toward the kinematics of far-side stars.  Since there
are $\sim 20$ stars between the peak and the median, we estimate this
bias as $\sim 20/241$ of the difference in mean proper motion of the
near-side and far-side subsamples, which we evaluate as $(\Delta
\mu_\ell,\Delta \mu_b) = (-0.02,0.02)\,\rm mas\,yr^{-1}$.  Since this
correction is an order of magnitude smaller than the observational
errors, we ignore it and adopt $\bv_{\bw,\perp}=0$.

We use equation (11) and apply the correction derived in equation (13)
to the values of the observed (relative) proper motion given in
equation (3) to obtain the absolute heliocentric proper motion
\be
(\mu_{\cl,l},\mu_{\cl,b}) = (1.4 \pm 0.2 ,-6.2 \pm
0.2)\,\rm mas\,yr^{-1}.
\ee

We note that the zero-point of the frame is uncertain by $0.17\,\rm
mas\, yr^{-1}$ in each direction due to the shot noise of the $\sim
330$ stars with scatter $\sim 3\,\rm mas\,yr^{-1}$ each.  Assuming that
the total solar motion is $\bv_{\odot}=(9,232,7)$ km s$^{-1}$ in the
($U,V,W$) reference frame ($U$ being positive inwards), we evaluate
equation (12), and find
\be
(v_{\cl,l},v_{\cl,b}) = (68 \pm 18 ,-208 \pm 23)\,\rm km\,s^{-1},
\ee 
where we have included the distance errors, but not included the
uncertainty in the overall distance scale.  Finally, we evaluate the
three space motion in ($U$,$V$,$W$) reference frame,
\be
\bv_\cl = (-30 \pm 7, 68 \pm 18, -208 \pm 23)\,\rm km\,s^{-1}.
\ee

That NGC 6522 has a significant motion away from the Galactic plane
indicates immediately that it is on a halo orbit (\cite{cud93}),
consistent with most other globular clusters which have [Fe/H] $< -1$
(\cite{zin85}; \cite{arm89}).  The cluster's current distance on the
Carney et al. (1995) scale ($7.1 \pm 0.3$ kpc) is very close to the
fixed Galactocentric distance of 7.8 kpc, so the cluster must be on an
orbit which takes it quite near ($< 1$ kpc) the center of the Galaxy.

We explore possible past orbits for NGC 6522 using our measurements of
the cluster's position and spatial motion.  For the inner Galaxy, we
assume an oblate logarithmic potential of the form
\be
\Phi(x,y,z) = {1\over 2}v_c^2 \ln \left[ x^2 + y^2 + (kz)^2 \right] + {\rm const.},
\ee
where $v_c$ is the circular velocity in the equatorial plane (taken as
220 km s$^{-1})$, and $k$ is a flattening term which we set in the
range $k = 1$ to 3.  We find that about $2 \times 10^6$ yr ago, the
cluster passed the Galactic center at a distance as close as $400 -
550$ pc, and that it generally does not achieve an apogalactic distance
in excess of 1500 pc.  Our calculation also shows that the time scale
for significant changes to the orbital energy through dynamical
friction is at least a few times $10^{10}$ yr.

Clusters that pass within a few hundred pc of the Galactic center are
likely to experience significant shocking from the time-variable
gravitational potential of the bulge and inner disk (most recently
\cite{gne97}; \cite{mur97}) and are unlikely to survive much longer
than another Hubble time.  NGC 6522 is also a core-collapsed cluster
(e.g., \cite{djk86}; \cite{lug95}) another sign that the cluster may
have experienced significant shocks (e.g., \cite{gne97} and references
therein).  We therefore suggest that NGC 6522 would be a good target
for detailed kinematic studies of high accuracy, to increase the number
of velocity-selected members and to obtain better estimates of the mass
and velocity dispersion.

\acknowledgments
Work by A.G.\ and P.P.\ were supported in part by grant AST 94-20746
from the NSF.  DT thanks Ruth C.\ Peterson for her helpful insights.

\clearpage

\figcaption[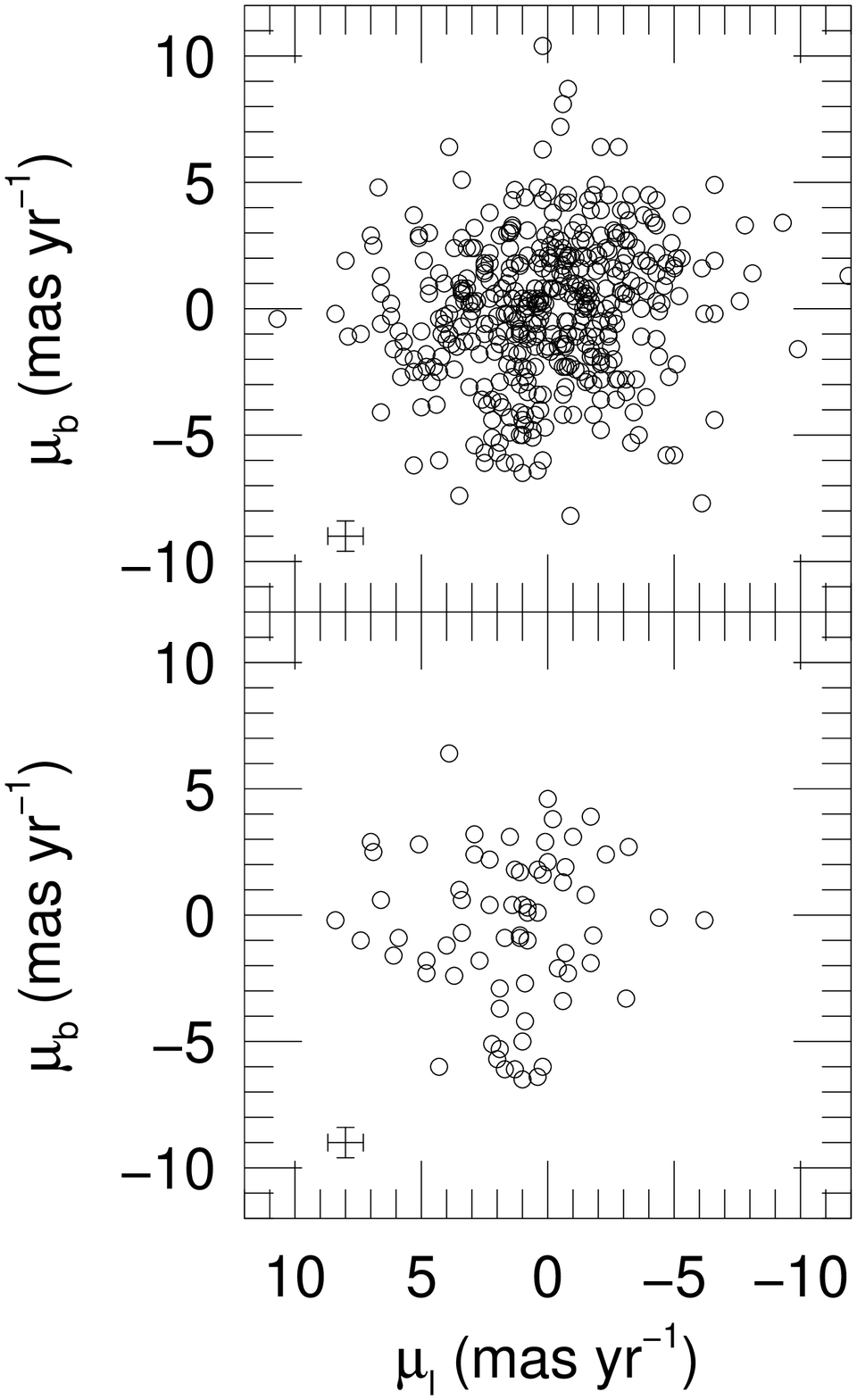]{
Proper motion vector point diagram for (top panel) all stars in the
Baade's Window survey (Jones et al. 1992) and (lower panel) those stars
located within $2\farcm 5$ of the cluster center and which have radial
velocities near that of NGC 6522. The units are milliarcsecond (mas)
per year. The error bars in the lower lefthand corner show the mean
error of measurement in $\mu_\ell$ and $\mu_b$, plotted as though they
were independent.  Because the proper motions in $\ell$ and $b$ were
obtained from the proper motion in right ascension and declination
through a coordinate rotation, the errors in $\mu_\ell$ and $\mu_b$ are
in fact correlated.
\label{fig1}} 

\figcaption[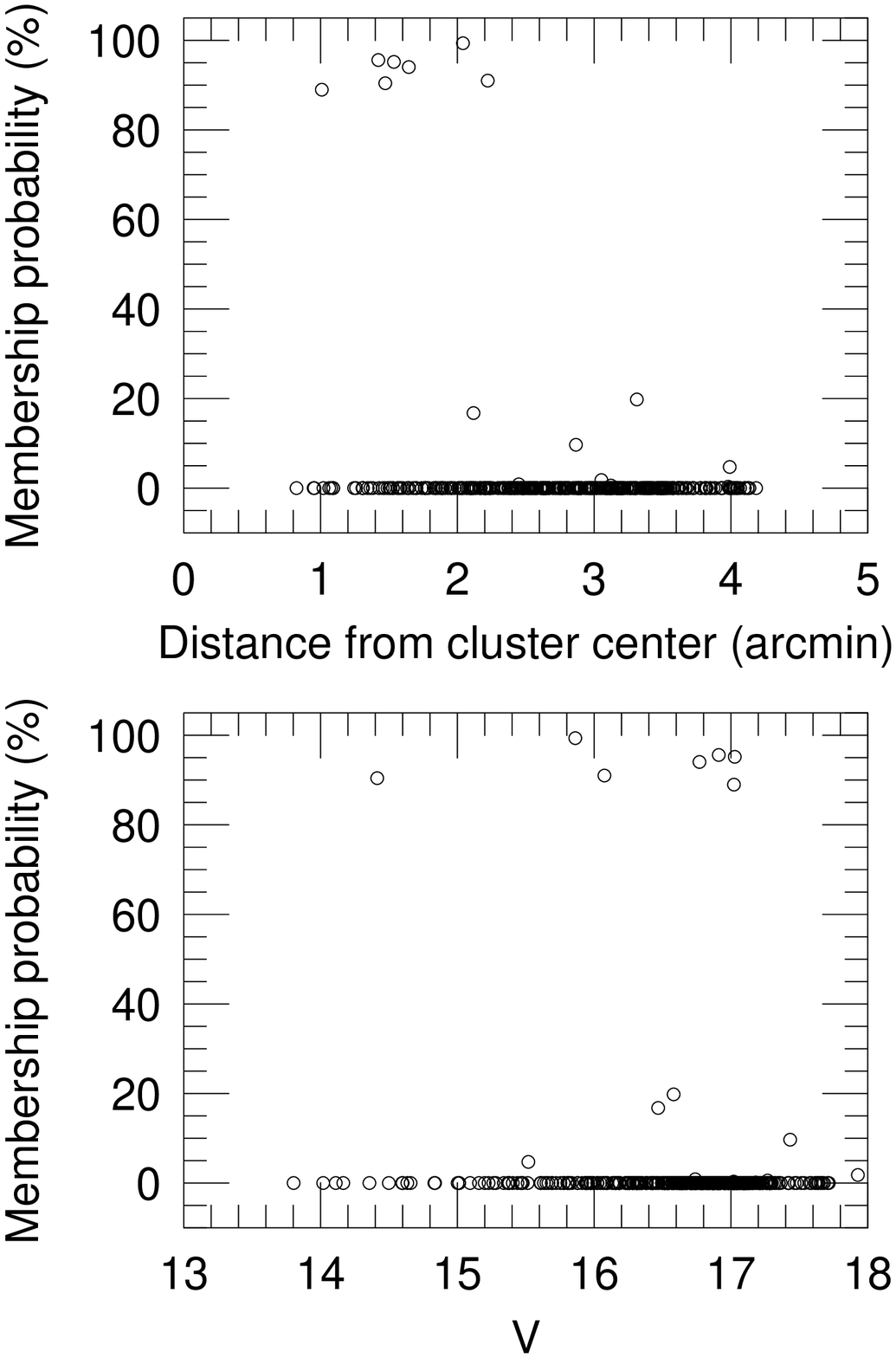]{
Membership probabilities in percent, as defined in the text, plotted
(top) against the distance from the center of NGC 6522, and (bottom)
against the $V$ magnitude.  The stars we identified as cluster members
have $P \geq 85\%$.
\label{fig2}}

\figcaption[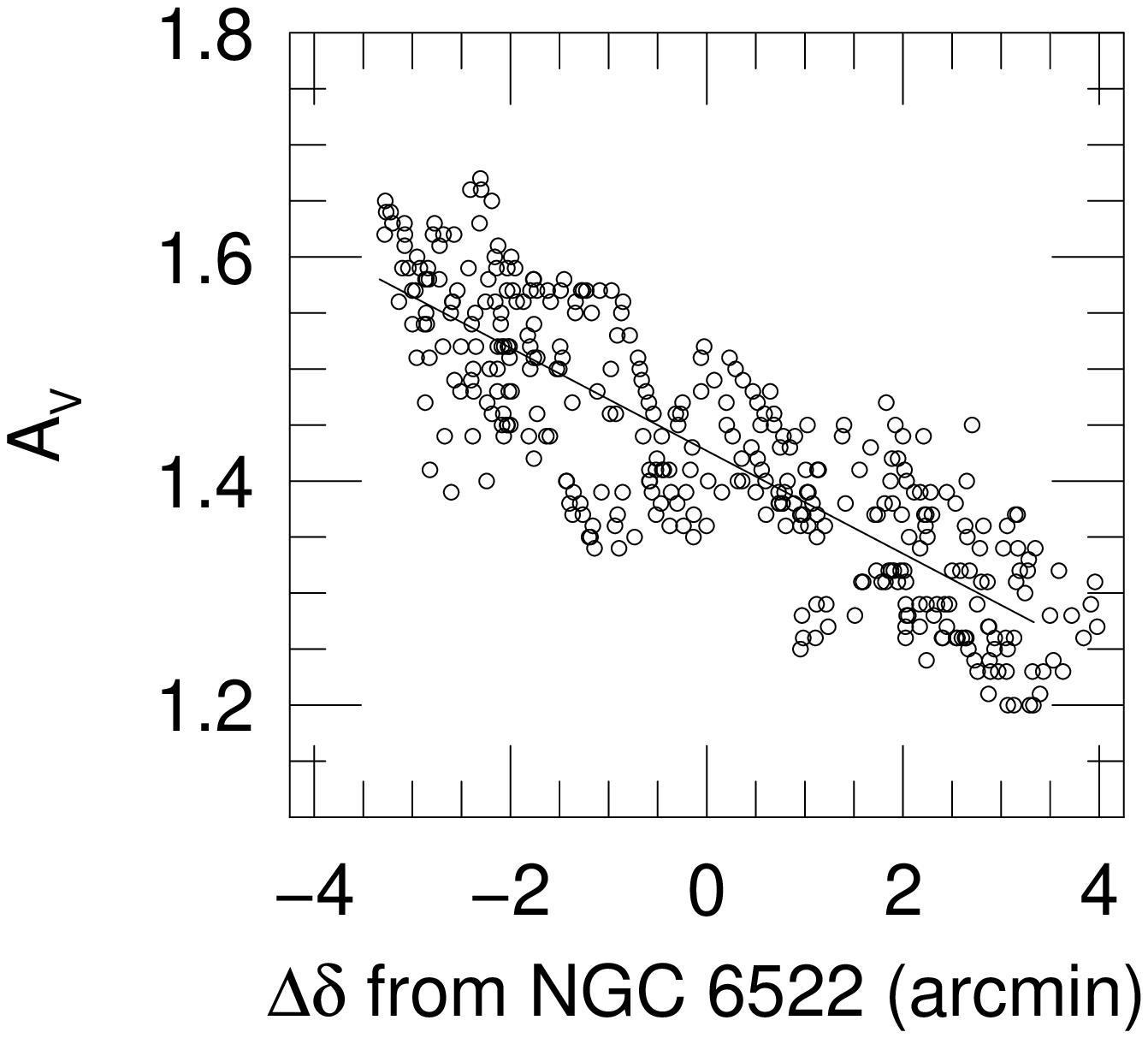]{
Visual extinction near NGC 6522.  Shown are the values of
the visual extinction $A_V$ for stars in the proper motion survey
as derived from the Stanek (1996) extinction map for Baade's Window,
plotted against the difference in declination between each star
and the center of NGC 6522.  The extinction values have been reduced
by $-0.10$ from the Stanek map, following the recalibration of the
zero point of the extinction in Gould et al. (1998).  The straight
line is a least-squares fit to the data, and was used to compute
reddenings within $2\arcmin$ of the cluster center, an area not
included in the Stanek (1996) extinction map.
\label{fig3}}

\figcaption[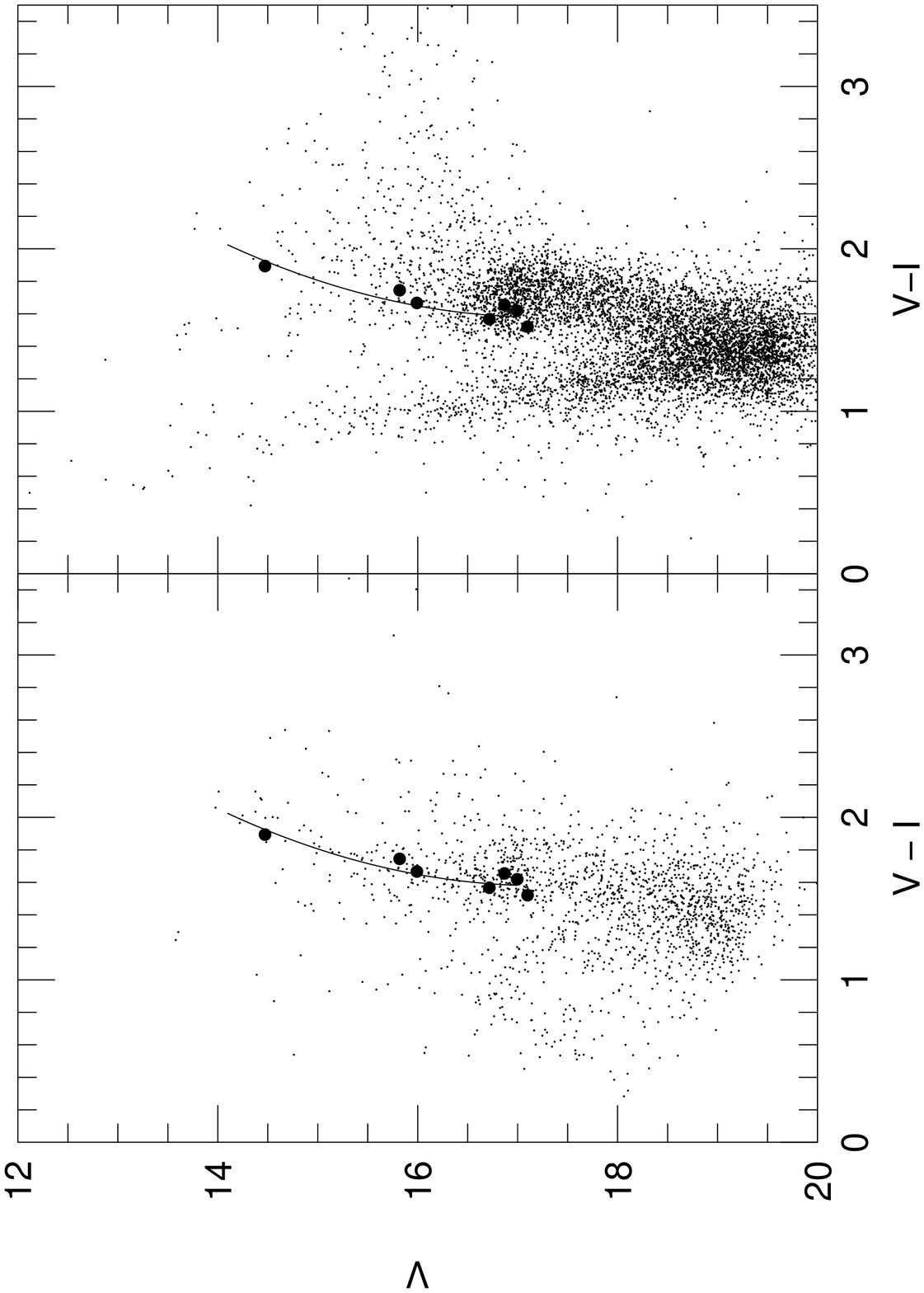]{
Color-magnitude diagrams in Baade's Window.  The left panel shows
photometry for stars within $1\farcm 5$ of NGC 6522, while the right
panel is for stars farther than $2\farcm 5$ from the cluster center.
The large filled points are for the probable cluster members in Table
1.  All photometry has been corrected to $A_V = 1.42$, a value
appropriate for the cluster center, as described in the text.  The
curved line is a quadratic fit to the cluster giant branch.
\label{fig4}}

\figcaption[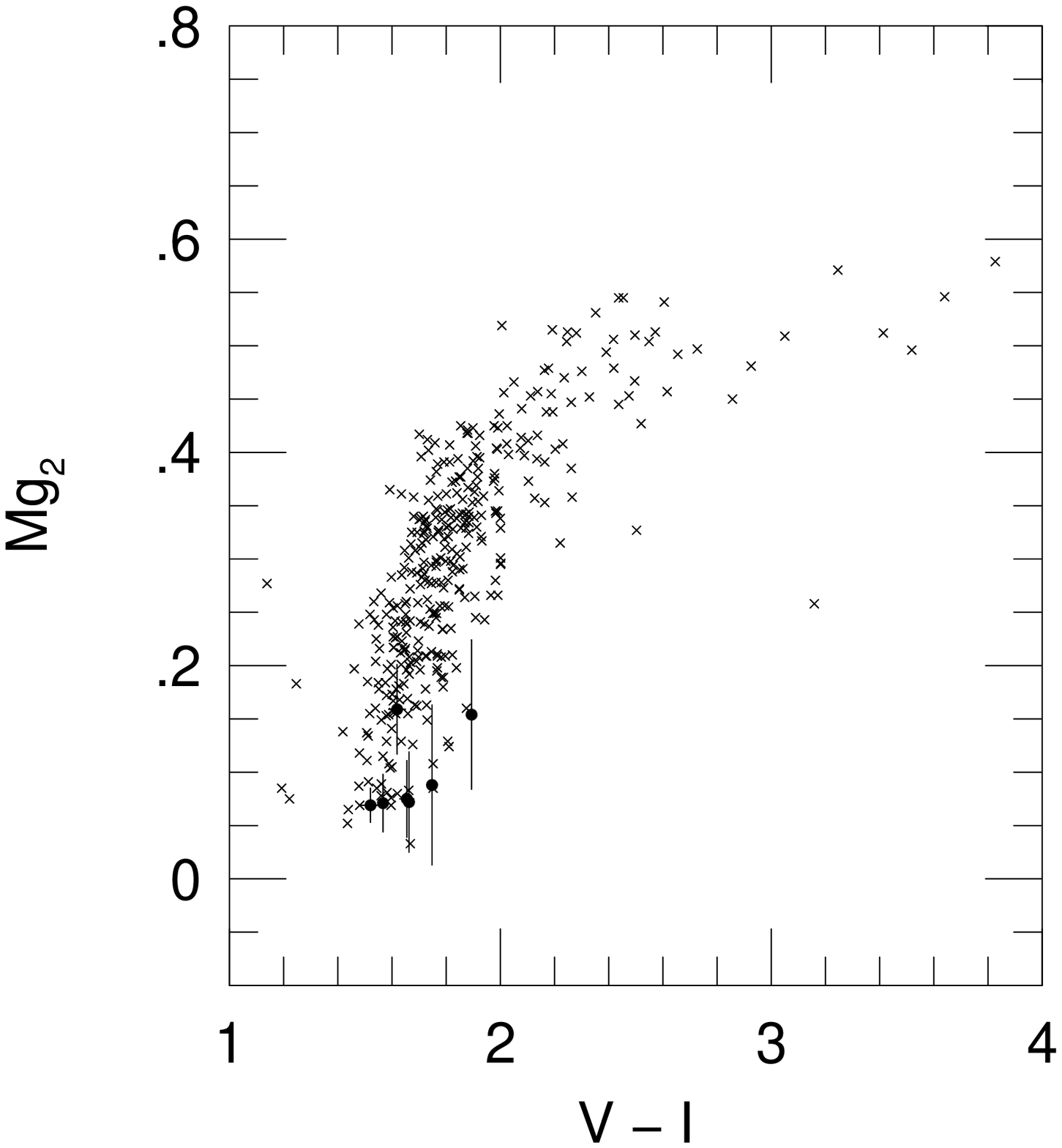]{
Mg$_2$ indices for SJW stars from Terndrup et al.\ (1995) as a function
of $V - I$ color.  The filled points with error bars are for the likely
members of NGC 6522 in Table 1, while the other points ($\times$) are
for the rest of the SJW sample.
\label{fig5}}

\clearpage


\clearpage
\begin{thebibliography}{}

\bibitem[Alcock et al. (1998a)]{alc98} Alcock, C., et al. 1998a, \apj,
494, 000 (astro-ph/9706292)

\bibitem[Alcock et al. 1998b]{alc98b} Alcock, C., et al. 1998b, \apj,
in preparation

\bibitem[Armandroff 1989]{arm89} Armandroff, T. E. 1989, \aj, 97, 375

\bibitem[Arp (1965)]{arp65} Arp, H, 1965, \apj, 141, 45

\bibitem[Barbuy et al. (1994)]{bar94} Barbuy, B., Ortolani, S., \&
Bica, E. 1994, \aap, 285, 871

\bibitem[Blanco 1984]{bla84} Blanco, B. M. 1984, \aj, 89, 1836

\bibitem[Carney et al. (1992)]{car92} Carney, B. W., Storm, J., \&
Jones, R. V. 1992, \apj, 386, 663

\bibitem[Carney et al. (1995)]{car95} Carney, B. W., Fulbright, J. P.,
Terndrup, D. M., Suntzeff, N. B., \& Walker, A. R. 1995, \aj, 110, 1674

\bibitem[Carretta \& Gratton (1997)]{cg97} Carretta, E., \& Gratton,
R.  G. 1997, \aaps, 121, 95

\bibitem[Castro et al. (1995)]{cas95} Castro, S., Barbuy B., Bica, E.,
Ortolani, S., \& Renzini A., \aaps, 111, 17

\bibitem[Cudworth \& Hanson 1993]{cud93} Cudworth, K. M., \& Hanson, R.
B. 1993, \aj, 105, 168

\bibitem[Da Costa \& Armandroff (1990)]{da90} Da Costa, G. S., \&
Armandroff, T. E. 1990, \aj, 100, 162

\bibitem[Dauphole et al.\ 1996]{dau96} Dauphole, B., Geffert, M.,
Colin, J., Ducourant, C., Odenkirchen, M., \& Tucholke, H.-J. 1996,
\aap, 313, 119

\bibitem[Dinescu et al. 1996]{din96} Dinescu, D. I., Girard, T. M., van
Altena, W. F., Yang, T.-G., \& Lee, Y.-W. 1996, \aj, 111, 1205

\bibitem[Djorgovski \& King 1986]{djk86} Djorgovski, S., \& King, I. R.
1986, \apj, 305, L61

\bibitem[Gnedin \& Ostriker 1997]{gne97} Gnedin, O. Y., \& Ostriker, J.
P. 1997, \apj, 474, 223

\bibitem[Gould et al. (1998)]{gould98} Gould, A., Popowski, P., \&
Terndrup, D. M. 1998, \apj, 492, 000 (LANL abstract astro-ph/9705020)

\bibitem[Harris 1996]{har96} Harris, W. E. 1996, \aj, 112, 1487

\bibitem[Ibata et al. 1997]{iba97} Ibata, R. A., Wyse, R. F., Gilmore,
G., Irwin, M. J., \& Suntzeff, N. B. 1997, \aj, 113, 634

\bibitem[Jones et al. 1994]{jon94} Jones, B. F., Klemola, A. R., \&
Lin, D. N. C. 1994, \aj, 107, 1333

\bibitem[Lee et al. (1990)]{ldz90} Lee, Y.-W., Demarque, P., \& Zinn,
R. 1990, \apj, 350, 155

\bibitem[Lugger et al. 1995]{lug95} Lugger, P. M., Cohn, H. N., \&
Grindlay, J. E. 1995, \apj, 439, 191

\bibitem[Murali \& Weinberg 1997]{mur97} Murali, C. \& Weinberg, M. D.
1997, \mnras, 288, 749

\bibitem[Pryor \& Meylan 1993]{pry93} Pryor, C., \& Meylan, G. 1993, in
Structure and Dynamics of Globular Clusters, ASP Conf. Series 50, ed.
G. Meylan and S. Djorgovski (ASP, San Francisco), p. 357

\bibitem[Rich (1990)]{ric90} Rich, R. M. 1990, \apj, 362, 604

\bibitem[Rich et al. 1998]{ric98} Rich, R. M., Terndrup, D. M., \&
Sadler, E. M. 1998, in preparation

\bibitem[Rutledge et al. (1997)]{rut97} Rutledge, G. A., Hesser, J. E.,
Stetson, P. B., Mateo, M., Simard, L., Bolte, M., Friel, E. D., \&
Copin, Y. 1997, \pasp, 109, 883

\bibitem[Sadler et al. 1996]{srt96} Sadler, E. M., Rich, R. M., \&
Terndrup, D. M. 1996, \aj, 112, 171

\bibitem[Sarajedini (1994)]{sar94} Sarajedini, A. 1994, \aj, 107, 618

\bibitem[Schweitzer \& Cudworth 1996]{sch96} Schweitzer, A. M., \&
Cudworth, K. M. 1996, in Formation of the Galactic Halo Inside and Out,
ASP Conf. Ser., 92, 532

\bibitem[Smith et al. (1976)]{smith76} Smith, M. G., Hesser, J. E., \&
Shawl, S. J. 1976, \apj, 206, 66

\bibitem[Sosin et al.\ (1997)]{sos97} Sosin, C., Rich, R. M., Piotto,
G., Djorgovski, S. G., King, I. R., Dorman, B., Liebert, J. \& Renzini,
A.  1997, in Abundances in Stellar Evolution, ed. R. T. Rood \& A.
Renzini (Cambridge Univ. Press:  Cambridge), in press

\bibitem[Spaenhauer et al.  1992]{spa92} Spaenhauer, A., Jones, B. F.,
\& Whitford, A. E. 1992, \aj, 103, 297 (SJW)

\bibitem[Stanek (1996)]{sta96} Stanek, K. Z. 1996, \apjl, 460, L37

\bibitem[Terndrup et al. 1995]{tsr95} Terndrup, D.M., Sadler, E.M., \&
Rich, R.M. 1995 \aj, 110, 1774

\bibitem[Terndrup \& Walker (1994)]{tw94} Terndrup, D. M., \& Walker,
A. R.  1994, \aj, 107, 1786

\bibitem[Walker \& Mack 1986]{wal86} Walker, A. R., \& Mack, P. 1986,
\mnras, 220, 69

\bibitem[Walker \& Terndrup 1991]{wal91} Walker, A. R., \& Terndrup,
D.  M.  1991, \apj, 378, 119

\bibitem[Webbink (1981)]{web81} Webbink, R. F. 1981, \apjs, 45, 259

\bibitem[Zinn (1983)]{zin83} Zinn, R. 1983, \apj, 293, 424

\bibitem[Zinn 1985]{zin85} Zinn, R. 1985, \apj, 293, 424

\bibitem[Zinn \& West (1984)]{zw84} Zinn, R., \& West, M. J. 1984,
\apjs, 55, 45

\end{thebibliography}
\end{document}